\documentclass[aps,prl,preprintnumbers,superscriptaddress,twocolumn,nofootinbib]{revtex4}
\usepackage[dvips]{graphicx}
\usepackage{bm,latexsym,amsmath,amssymb,amsfonts,mathrsfs}
\usepackage{color}
\usepackage[normalem]{ulem}
\input{colordvi.tex}

\begin{document}

\preprint{KOBE-COSMO-19-12}
\title{String Regge trajectory in de Sitter space and implications for inflation}

\author{Toshifumi Noumi}
\affiliation{Department of Physics, Kobe University, Kobe 657-8501, Japan}

\author{Toshiaki Takeuchi}
\affiliation{Department of Physics, Kobe University, Kobe 657-8501, Japan}

\author{Siyi Zhou}
\affiliation{Department of Physics and Jockey Club Institute for Advanced Study,
Hong Kong University of Science and Technology, Hong Kong}
\affiliation{The Oskar Klein Centre for Cosmoparticle Physics and Department of Physics, Stockholm University, AlbaNova, 106 91 Stockholm, Sweden.}

\begin{abstract}

We study the spectrum of semiclassical rotating strings in de Sitter space and its consistency. Even though a naive extrapolation of the linear Regge trajectory on flat space implies a violation of the Higuchi bound (a unitarity bound on the mass of higher-spin particles in de Sitter space), the curved space effects turn out to modify the trajectory to respect the bound. Interestingly, as a consequence of accelerated expansion, there exists a maximum spin for each Regge trajectory, which is helpful to make the spectrum consistent with the Higuchi bound, but at the same time, it could be an obstruction to stringy UV completion based on an infinite higher-spin tower. By pushing further this observation, we demonstrate that the vacuum energy $V$ inflating the universe has to be bounded by the string scale $M_s$ as $V\lesssim M_s^4$, if UV completion is achieved with the leading Regge trajectory of higher spin states up to the 4D Planck scale. Its application to inflation in the early universe implies an upper bound on the tensor-to-scalar ratio, $r\lesssim 0.01\times(M_s/10^{16} \text{GeV})^{4}$, which is within the scope of the near future CMB experiments. We also discuss another possibility that UV completion is achieved by multiple Regge trajectories.

\end{abstract}

\maketitle

\section{Introduction}

Our universe has experienced accelerated expansions both in the early universe and at present. For string theory to describe our real world, these accelerated expansions have to be accommodated in a consistent manner. The pioneering work in this direction is the KKLT scenario~\cite{Kachru:2003aw}, which proposed a concrete realization of de Sitter (dS) space in string theory by evading assumptions of the Maldacena-Nunez no-go theorem~\cite{Maldacena:2000mw}\cite{footnote1}. While the proposal has passed various nontrivial consistency checks, it relies on non-perturbative effects, which are not fully understood yet and have been studied intensively (see, e.g.,~\cite{Danielsson:2018ztv} for a review). In this paper, as a complementary approach to de Sitter space in string theory, we would like to discuss the semiclassical spectrum of would-be string theory on de Sitter space and its consistency. We also explore implications to inflation at the early universe.

\subsubsection{String Regge trajectory vs. the Higuchi bound}

\begin{figure}[t] 
	\centering 
	\includegraphics[width=6cm, bb=0 0 450 444]{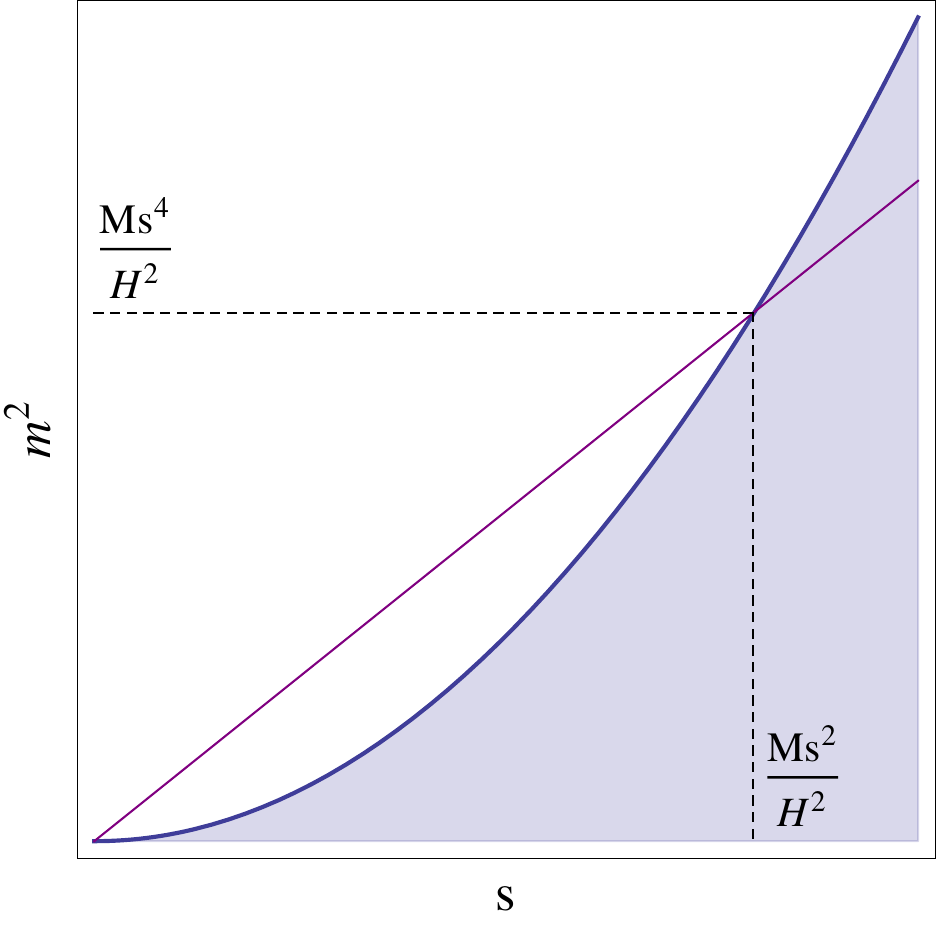}  
	\caption{Higuchi bound vs linear Regge trajectory: The shaded region is prohibited by the Higuchi bound \eqref{Higuchi}. The linear Regge trajectory (the straight line) enters the prohibited region at a critical spin $s_c\simeq M_s^2/H^2$.} \label{linearregge}
\end{figure}

We begin our argument by a simple observation that string theory might potentially be in conflict with the so-called Higuchi bound~\cite{Higuchi:1986py},
\begin{align}
\label{Higuchi}
m^2\geq s(s-1)H^2\,,
\end{align}
on the mass of higher-spin particles in de Sitter space~\cite{footnote2}.
Here $m$ and $s$ are the mass and spin of a particle and $H=1/R$ is the Hubble parameter with $R$ being the de Sitter radius. Massive higher-spin particles violating the Higuchi bound contain helicity modes with a negative norm, hence they are prohibited by unitarity. Immediately, we find that a linear Regge trajectory,
\begin{align}
m^2\simeq sM_s^2
\quad
\text{for large $s$},
\end{align}
violates the bound~\eqref{Higuchi} beyond the critical mass $m_c$ and spin $s_c$ given by (see Fig.~\ref{linearregge})
\begin{align}
m_c\simeq \frac{M_s^2}{H}\,,
\quad
s_c\simeq \frac{M_s^2}{H^2}\,.
\end{align}
Here $M_s$ is identified with the string scale {\it if} we extrapolate the leading Regge trajectory of the flat space string. Then, one might wonder that perturbative string is in conflict with the Higuchi bound and therefore it cannot be realized on dS consistently. However, it is too quick to conclude because a typical length of the string at the critical value is near the Hubble horizon scale:
\begin{align}
\ell\sim \frac{m_c}{M_s}\ell_s\sim H^{-1}\,,
\end{align}
where we introduced $\ell_s\sim1/M_s$. In this regime, the linear Regge trajectory may be modified by the spacetime curvature effects. In order for perturbative string on de Sitter space to be consistent, the string spectrum has to be modified at least to respect the Higuchi bound. In the first half of the paper, we demonstrate that the curved space effects indeed modify the Regge trajectory appropriately to respect the Higuchi bound.

\subsubsection{Upper bound on inflation scale}

Interestingly, the modified Regge trajectory accommodates a maximum spin due to accelerated cosmic expansion~\cite{deVega:1996mv}.
On one hand, this is helpful to make the string spectrum consistent with the Higuchi bound as illustrated in Fig.\ref{reggebehaviorsl}. However, let us recall that existence of an {\it infinite} higher-spin tower (the Regge tower) is crucial in string theory to make mild the high-energy scattering and UV complete gravity in a weakly coupled regime. It is therefore highly nontrivial for the modified spectrum to maintain the mildness of high-energy string scattering.

In the latter half of the paper, we discuss implications of such high-energy scattering.
In string theory on flat space and anti-de Sitter (AdS) space, the leading Regge trajectory contains infinitely many higher spin states, and these higher spin states are crucial for UV completion.
Assuming the same scenario on de Sitter space,
we postulate that there exist sufficiently many higher-spin particles up to some UV scale $\Lambda$, which would make mild high-energy scattering in the regime $M_s<E<\Lambda$. If we choose this scale to be the 4D Planck scale $\Lambda\sim M_{\rm Pl}$,
the condition translates into an upper bound on the vacuum energy, $V\lesssim M_s^4$. Its application to inflation at the early universe implies an upper bound on the tensor-to-scalar ratio, $r\lesssim 0.01\times(M_s/10^{16} \text{GeV})^{4}$. For the typical value of the string scale $M_s\sim 10^{16} \text{GeV}$,
we arrive at $r\lesssim0.01$ under the assumptions, whose boundary is within the scope of the near future CMB experiments!

\section{Regge trajectory on de Sitter}

In this section we study semiclassical strings rotating around the origin of static coordinates of de Sitter space (see~\cite{deVega:1996mv} for pioneering works on Regge trajectories on various curved backgrounds),
\begin{align}
ds^2=R^2\left[-(1-r^2)dt^2+\frac{dr^2}{1-r^2}+r^2d\Omega_{2}^2
\right]
\,,
\end{align}
where $R$ is the de Sitter radius, $d\Omega_{2}^2$ is the line element on a unit sphere $S^{2}$, and $0\leq r<1$. The cosmological horizon of the static observer sitting at the origin $r=0$ is located at $r=1$. For our purpose, it is convenient to introduce $r=\sin \rho$, in terms of which the metric reads
\begin{align}
ds^2=R^2\Big[-\cos^2\rho \,dt^2+d\rho^2+\sin^2\rho\,d\Omega_{2}^2
\Big]
\,.
\end{align}
In these coordinates, the observer and the cosmological horizon are located at $\rho=0$ and $\rho=\pi/2$, respectively. Also note that global coordinates on anti-de Sitter space can be obtained by a Wick rotation,
\begin{align}
\rho\to -i\rho\,,\quad
t\to it\,,
\quad
R^2\to -R^2\,.
\end{align}
As it suggests, we may generalize semiclassical strings on  anti-de Sitter (AdS) space used for the precision test of the AdS/CFT correspondence (see, e.g.,~\cite{Tseytlin:2010jv} for a review) to dS in a straightforward manner. Similarly to the AdS case, we assume that the string scale $M_s\sim \alpha'^{-1/2}$ is sufficiently higher than the Hubble scale $H=1/R$ for validity of the semiclassical approximation.

The present paper focuses on folded strings, whose AdS counterpart is known as the Gubser-Klebanov-Polyakov (GKP) string~\cite{Gubser:2002tv} originally constructed by de Vega and Egusquiza~\cite{deVega:1996mv}. More general solutions will be studied in the future work.

\begin{figure}[t] 
	\centering 
	\includegraphics[width=5cm, bb=0 0 216 214]{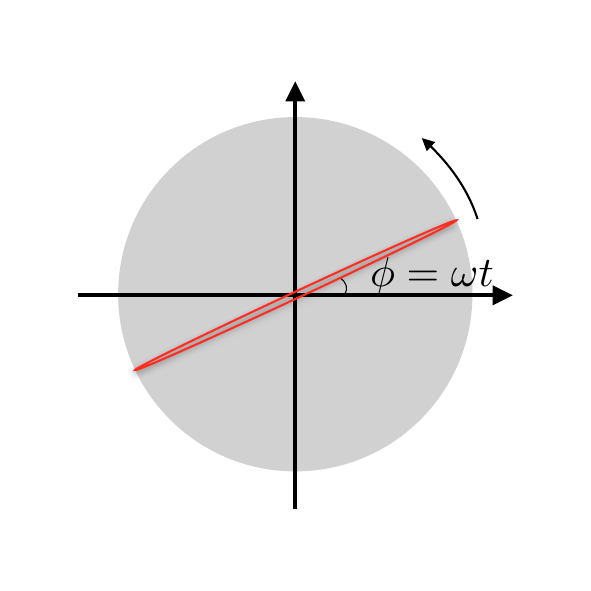}  
	\caption{A folded closed string rotating around the origin $\rho=0$ along an equator of $S^{2}$ with angular velocity $\omega$.} \label{rotation}
\end{figure}

\subsection{Semiclassical rotating strings}

Suppose that a folded closed string is rotating around the origin along an equator of $S^{2}$:
\begin{align}
\phi=\omega t\,,
\end{align}
where $\phi$ represents the angle of the equator and $\omega$ is the angular velocity (see Fig.~\ref{rotation}). If we use the gauge $\tau=t$ in the Nambu-Goto action, the string Lagrangian reads~\cite{footnote3}
\begin{align}
L=-4\frac{R^2}{2\pi\alpha'}\int_0^{\rho_0}d\rho\sqrt{\cos^2\rho-\omega^2\sin^2\rho}\,,
\end{align}
where $\rho$ is a function of $\sigma$.
The centrifugal force stretches the string to a distance $\rho_0$ determined by
\begin{align}
\cot^2\rho_0=\omega^2\,.
\end{align}
This is in turn the maximum distance allowed by causality of the Nambu-Goto action. Strings stretching beyond this would imply super-luminal velocities.

\begin{figure}[t] 
	\centering 
	\includegraphics[width=6cm, bb=0 0 450 442]{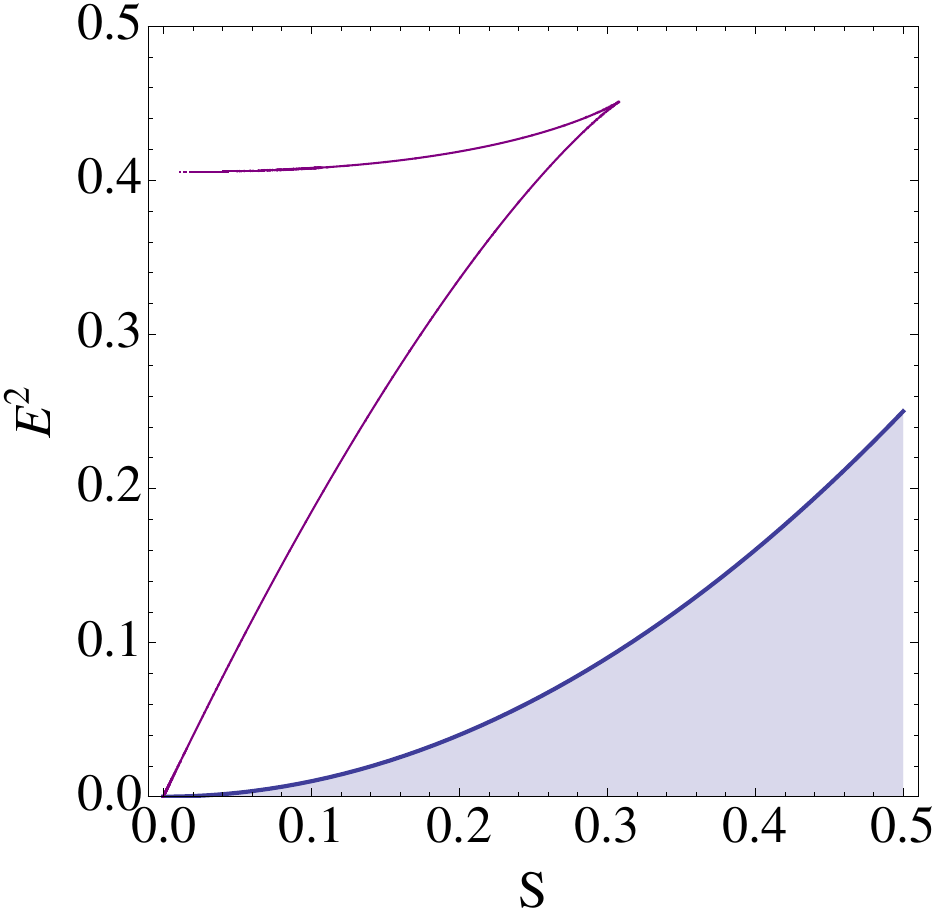}  
	\caption{Leading Regge trajectory vs. Higuchi bound: The leading Regge trajectory (the spiky curve) turns back to lower spins at a maximum spin before hitting the shaded region prohibited by the Higuchi bound. The spin $S$ and the energy $E$ are plotted in the units of $R^2/\alpha'$ and $R/\alpha'$, respectively. The same units are used in Fig.~\ref{ES_rho} and Fig.~\ref{reggebehavior}.} \label{reggebehaviorsl}
\end{figure}

\begin{figure}[t] 
	\centering 
	\includegraphics[width=6.5cm]{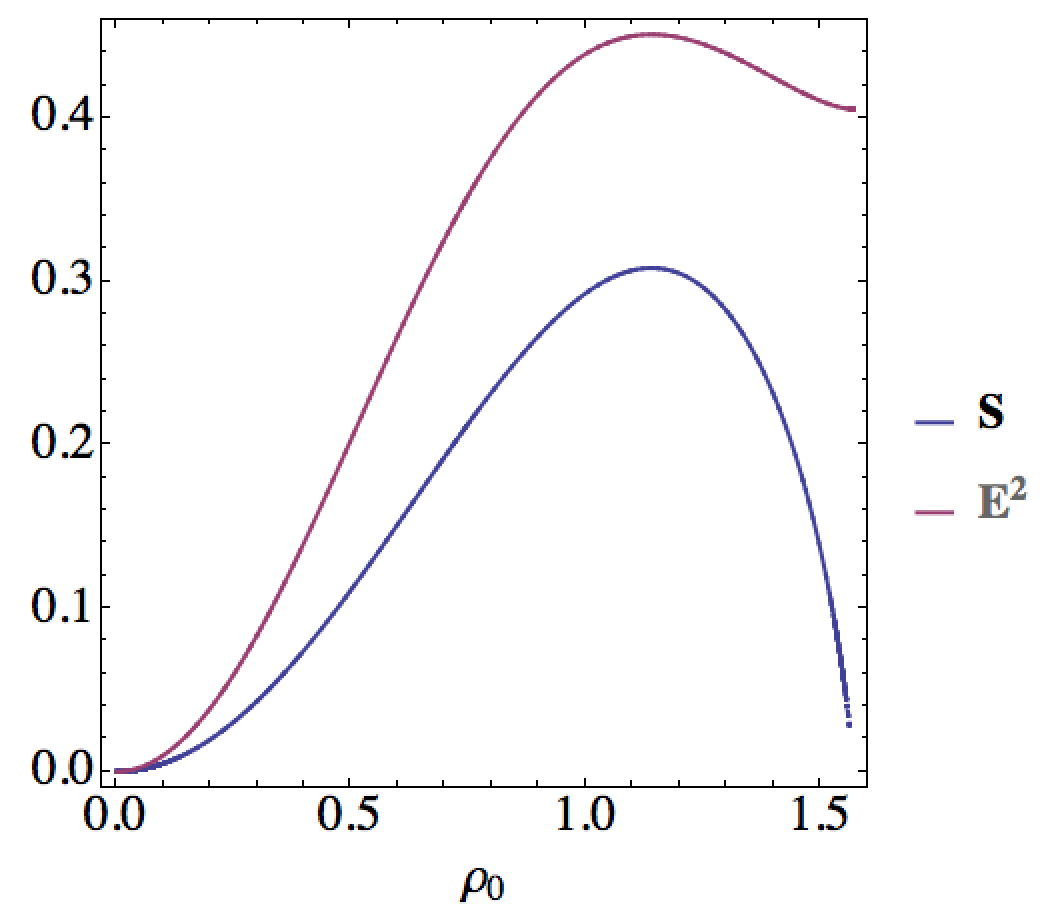}  
	\caption{The spin $S$ and energy squared $E^2$ as functions of $\rho_0$:
	We find that both have a peak at $\rho_0\simeq1.14$. Recall that there are two sources of the energy: one proportional to the string length and the other from the string rotation. Then, the maximum energy appears when the spin takes the maximum value. As a result, Fig.~\ref{reggebehaviorsl} has a spiky shape.
	} \label{ES_rho}
\end{figure}

Similarly, the energy $E$ (dual to $Rt$) and spin $S$ read
\begin{align}
\label{1_fold_E}
E&=4\frac{R}{2\pi\alpha'}\int_0^{\rho_0}d\rho\frac{\cos^2\rho}{\sqrt{\cos^2\rho-\omega^2\sin^2\rho}}\,,
\\
\label{1_fold_S}
S&=4\frac{R^2}{2\pi\alpha'}\int_0^{\rho_0}d\rho\frac{\omega\sin^2\rho}{\sqrt{\cos^2\rho-\omega^2\sin^2\rho}}\,.
\end{align}
In terms of incomplete elliptic integrals,
\begin{align}
\mathcal{E}(\varphi|k^2)&=\int_0^\varphi d\theta\sqrt{1-k^2\sin^2\theta}\,,
\\
\mathcal{F}(\varphi|k^2)&=\int_0^\varphi d\theta\frac{1}{\sqrt{1-k^2\sin^2\theta}}\,,
\end{align}
we may write these integrals as
\begin{align}
\nonumber
E&=\frac{R}{\pi\alpha'}\Big[\mathcal{E}(\rho_0| \csc^2\!\rho_0) + \cot^2\!\rho_0 \mathcal{F}(\rho_0| \csc^2\!\rho_0)\Big]
\\
\label{energy}
&\quad
\times(1-\cos2\rho_0)
 \,,
\\\label{spin}
S&=\frac{R^2}{\pi\alpha'}\Big[\!-\!\mathcal{E}(\rho_0| \csc^2\!\rho_0) +  \mathcal{F}(\rho_0| \csc^2\!\rho_0)\Big]\sin2\rho_0\,.
\end{align}
This provides the energy-spin relation through the parameter $\rho_0$ characterizing the length of the string. In Fig.~\ref{reggebehaviorsl} we plot the energy squared $E^2$ as a function of the spin $S$. See also Fig.~\ref{ES_rho} for $S$ and $E^2$ as functions of $\rho_0$.
As we explain below, there exists a maximum spin at the intermediate scale~\cite{deVega:1996mv}\cite{footnote0}. The leading Regge trajectory then turns out to be consistent with the Higuchi bound.

\subsection{Short strings}

Let us first look at the spectrum of short strings. When the angular velocity $\omega$ is large, strings cannot be so long because of causality. In this regime, we have $\rho_0\simeq\omega^{-1}$ and the string does not feel the spacetime curvature. The energy and spin are then the same as the flat space ones,
\begin{align}
E\simeq \frac{R}{\alpha'}\rho_0\,,
\quad
S\simeq \frac{R^2}{2\alpha'}\rho_0^2\,,
\end{align}
which enjoy the linear Regge trajectory:
\begin{align}
E^2\simeq\frac{2}{\alpha'}S\,.
\end{align}

\subsection{Long strings}

Another extremal case is the small $\omega$ limit, under which we have $\rho_0\simeq\pi/2-\omega$. In this regime, the string end points approach to the cosmological horizon $\rho=\pi/2$, so that the spacetime curvature is not negligible. It is easy to evaluate the energy and spin as
\begin{align}
E\simeq\frac{2R}{\pi\alpha'}\,,
\quad
S\simeq-\frac{2R^2}{\pi\alpha'}\omega\ln\omega\,.
\end{align}
Interestingly, the spin vanishes in this limit, whereas the mass approaches to a fixed value. This is in a sharp contrast to the AdS case~\cite{deVega:1996mv,Gubser:2002tv}.

It will be instructive to elaborate on the qualitative difference of Regge trajectories on flat space, AdS, and dS. As we mentioned, the length of rotating strings characterized by $\rho_0$ is determined by causality of the Nambu-Goto string. On flat space, the velocity of the string end points is given by $\rho_0\omega$, so that causality tells us that $\rho_0=\omega^{-1}$. In particular, $\omega$ and $\rho_0$ can take an arbitrary positive value. On the other hand, on AdS, there is a lower-bound $\omega>1$ on $\omega$, which is saturated by rotating strings touching the AdS boundary, essentially because AdS is compact. In both cases, the spin increases as we decrease the angular velocity $\omega$ (or equivalently as the string length $\rho_0$ increases).

Finally, let us consider the de Sitter case, where the accelerated expansion of the universe plays a crucial role. First, the Hubble law implies that velocity exceeds the speed of light beyond the Hubble horizon. Therefore, the end points of a folded closed string cannot stretch beyond the horizon. Note that when the end points touch the cosmological horizon, their velocity coincides with the speed of light for $\omega=0$, so that any nonzero $\omega$ leads to a causality violation. Therefore, there exists a maximum value of the string length $\rho_0$, for which the angular velocity $\omega$ and then the spin $S$ have to vanish.

In this way, the spectrum of long strings on de Sitter is qualitatively different from the flat space and AdS ones. In particular, the longest string has a vanishing spin and a finite mass due to the accelerated expansion.

\subsection{Maximum spin}

We have argued that both the shortest string $\rho_0=0$ and the longest one $\rho_0=\pi/2$ have a vanishing spin $S=0$. It suggests that there exists a maximum spin $S_*$ on the Regge trajectory. From the expression~\eqref{spin}, we find that the maximum spin appears at $\rho_0 \simeq 1.14$, for which the mass $E_*$ and the spin $S_*$ are
\begin{align}
\label{maximum_spin}
E_*\simeq 0.67\times R/\alpha'\,,
\quad
S_*\simeq 0.31\times R^2/\alpha'\,,
\end{align}
which is above the Higuchi bound. The full trajectory for $0\leq \rho_0\leq\pi/2$ is give in Fig.~\ref{reggebehaviorsl}, which shows that the semiclassical rotating strings have a spectrum consistent with the Higuchi bound. Even though our analysis focused on a rotating folded closed string, any more internal structures will increase the mass, hence the spectrum shown in Fig.~\ref{reggebehaviorsl} will provide the leading Regge trajectory. We therefore conclude that the semiclassical string spectrum on de Sitter space is consistent with the Higuchi bound. This is one of our main results in this paper. Note that our conclusion is independent of the ratio $M_s/H$ as long as $M_s\gg H$ (i.e., within the validity of the semiclassical approximation).

\section{High energy scattering}

We have shown that semiclassical rotating strings have a spectrum consistent with the Higuchi bound. Does it mean that perturbative string theory can in principle be realized on de Sitter space consistently? Of course, it is not the end of the story. In string theory, existence of an infinite higher spin tower (the Regge tower) is crucial to make mild the high-energy behavior of scattering amplitudes and to UV complete gravity in a weakly coupled regime. Therefore, the existence of a maximum spin in the Regge trajectory makes it nontrivial to maintain the mildness of high-energy scattering. In this section we would like to explore two possibilities.

\subsection{UV completion by the leading Regge trajectory}

First we consider a possibility that UV completion is achieved with higher-spin states on the leading Regge trajectory, just like the flat space and AdS case. In order to UV complete gravity in a weakly coupled regime, we would need sufficiently many higher-spin states from the string scale $M_s$ up to some UV scale $\Lambda$, which make mild high-energy scattering in the regime $M_s<E<\Lambda$.
For this to happen, the mass $E_*\sim R/\alpha'\sim M_s^2/H$ of the maximum spin state in the leading Regge trajectory has to be bigger than the UV scale $\Lambda$:
\begin{align}
E_*\gtrsim 
\Lambda
\,,
\end{align}
which can be rephrased as
\begin{align}
H \lesssim \frac{M_s^2}{\Lambda}\,.
\end{align}
Interestingly, if we take the scale $\Lambda$ to be the 4D Planck scale $\Lambda\sim M_{\rm Pl}$~\cite{footnote4}, this condition provides an upper bound on the vacuum energy inflating the universe as
\begin{align}
\label{bound_on_V}
V =3M_{\rm Pl}^2H^2 \lesssim M_s^4\,.
\end{align}
Applying the bound to the inflation at the early universe gives an upper bound on the tensor-to-scalar ratio:
\begin{align}
\label{bound_on_r}
r=0.01\times \frac{V}{(10^{16} \text{GeV})^4}\lesssim 0.01\times\left(\frac{M_s}{10^{16} \text{GeV}}\right)^{4}\,,
\end{align}
where we assumed single-field inflation with a canonical kinetic term.
If the string scale is around $10^{16}$ GeV or below, the bound saturates the target sensitivity of the near future observations of CMB $B$-modes such as the LiteBIRD experiment~\cite{Hazumi:2019lys}. {\it If} the primordial gravitational waves were not detected in the near future, such a theoretical bound could be behind as an obstruction to high-scale inflation.

\begin{figure}[t] 
	\centering 
	\includegraphics[width=6cm, bb=0 0 450 468]{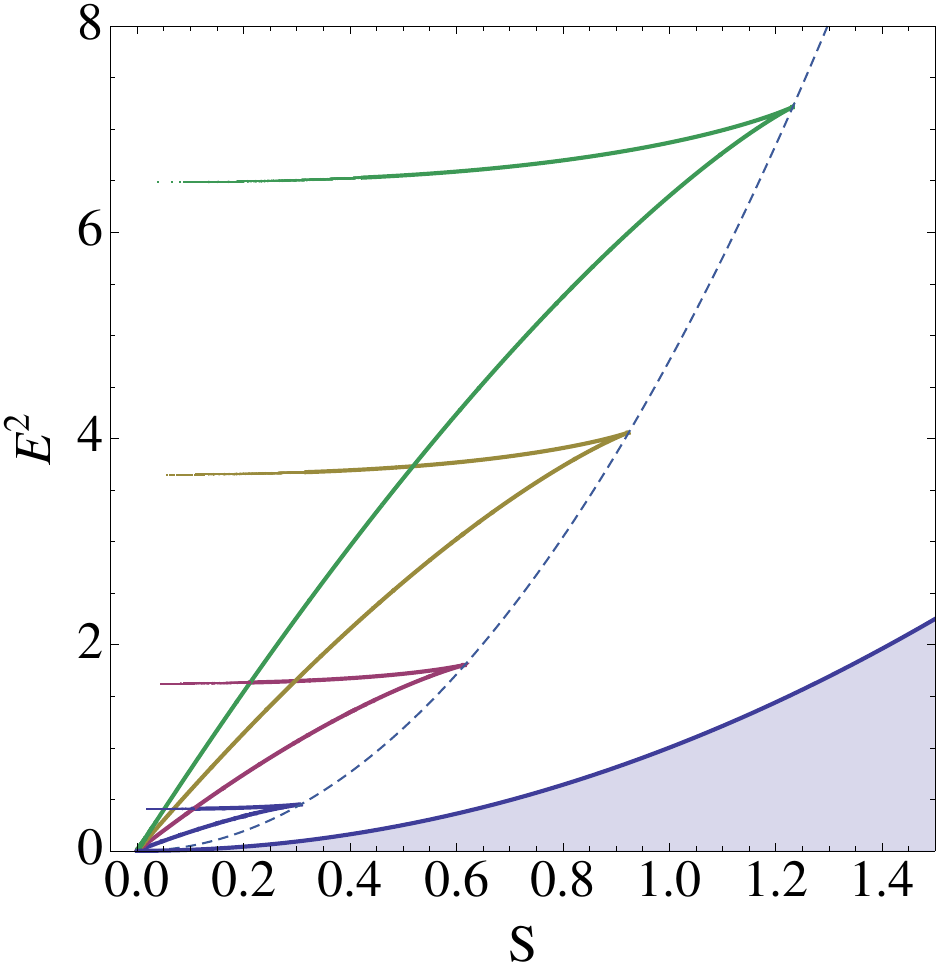}  
	\caption{Multiple Regge trajectories from $N$-folded closed strings: The spiky curves are Regge trajectories for $N=1,2,3,4$ from the bottom and the dotted curve is a quadratic curve on which the maximum spin state for each trajectory is located. The shaded region is prohibited by the Higuchi bound as before.} \label{reggebehavior}
\end{figure}

\subsection{UV completion by multiple Regge trajectories}

While it is very exciting if there exists a universal upper bound on the tensor-to-scalar ratio of the form~\eqref{bound_on_r}, there is another possibility of UV completion, namely higher-spin states from multiple Regge trajectories might help to make mild the amplitudes. To be concrete, as an illustrative example, let us consider a rotating closed string with $N$-folds, whose total length is $4N\rho_0$. The energy and spin of such $N$-folded closed strings are simply
\begin{align}
E_N&=NE\,,\quad S_N=NS\,,
\end{align}
where $E$ and $S$ are the energy and spin of the one-fold closed string~\eqref{1_fold_E}-\eqref{1_fold_S}. In the short string regime, $N\geq2$ provides subleading Regge trajectories as
\begin{align}
E^2_N=N\times \frac{2S_N}{\alpha'}\,.
\end{align}
Also, the mass $E_{N*}$ and spin $S_{N*}$ of the maximum spin state of each trajectory read
\begin{align}
(E_{N*}^2,S_{N*})=(N^2E_*,NS_*)\,,
\end{align}
where $E_*$ and $S_*$ are given in Eq.~\eqref{maximum_spin}. As depicted in Fig.~\ref{reggebehavior}, each Regge trajectory has a maximum spin state before touching the Higuchi bound and then the trajectory turns back into lower spins. In particular, we find an infinite higher-spin tower with an increasing spin above the Higuchi bound. {\it If the  interactions are appropriately arranged,} it would in principle be possible that these infinite higher-spin states make mild high-energy scattering beyond the energy scale $E=E_*$. If it is indeed the case, the bound~\eqref{bound_on_V} on the vacuum energy and therefore that on the tensor-to-scalar ratio~\eqref{bound_on_r} can be evaded, so that there remains a room for high-scale inflation. The exploration of such scattering amplitudes are beyond our scope in the present paper, leaving it for future work.

\section{Outlook}

In this paper we studied the semiclassical spectrum of a folded closed string rotating in de Sitter space. In a sharp contrast to Regge trajectories on flat space and AdS, the dS trajectory has a maximum spin because of the cosmic horizon, which is crucial to evade a potential conflict with the Higuchi bound. We also discussed implications to high-energy scattering. Due to existence of the maximum spin in each trajectory, it is nontrivial to make mild the high-energy behavior of scattering amplitudes. We demonstrated that an interesting upper bound on the vacuum energy $V\lesssim M_s^4$ is implied if we assume that UV completion is achieved with the leading Regge trajectory of higher spin states up to the 4D Planck scale.
To clarify how universal this bound is, it is necessary, e.g., to explore if high-energy scattering can be UV completed with multiple Regge trajectories each of which has a finite number of higher-spin states.

There are various future directions to explore along the line of the present work. One possible direction is to generalize our analysis to other semiclassical strings. For example, we expect that existence of a maximum spin in each Regge trajectory is a universal nature of strings on de Sitter space. It would be interesting to see it and clarify how the string spectrum recovers the Higuchi bound in more general. A more challenging, but important direction is a generalization to higher-point correlations. In the past several years, an impressive progress has been made in applications of the world-sheet integrability to holographic correlation functions~\cite{Zarembo:2010rr,Costa:2010rz,Janik:2011bd,Kazama:2011cp,Caetano:2012ac,Kazama:2013qsa,Kazama:2016cfl}. It would be interesting to generalize these developments to the world-sheet theory on de Sitter space. Such a direction will provide a fundamental tool to study string scattering on de Sitter space, which is crucial to discuss consistency of perturbative string on de Sitter space. We hope to revisit these issues elsewhere in the near future.

\begin{acknowledgments}

\bigskip
\noindent
{\bf Note added:} While we are finalizing the draft, we have learned that a potential conflict of the string Regge trajectory and the Higuchi bound was mentioned at the conference ``String Phenomenology 2019"~\cite{LPV,LPV2}. The argument there was purely based on the linear Regge trajectory on flat space and equivalent to our observation presented in the introduction. However, one of our main points is that the Regge trajectory is modified by the Hubble horizon effects to respect the Higuchi bound and there exists a maximum spin for each trajectory, which is different from~\cite{LPV,LPV2} based on the flat space spectrum.

\section{Acknowledgments}
We would like to thank Stefano Andriolo, Shinji Hirano, Hiroshi Isono, Mitsuhiro Kato, Suro Kim, Shota Komatsu, Yuji Okawa, Norma Sanchez, Jiro Soda, Pablo Soler and Daisuke Yoshida for valuable discussion and comments on the draft.
TN is supported in part by JSPS KAKENHI Grant Numbers JP17H02894, JP18K13539, and 20H01902 and 
MEXT KAKENHI Grant Number JP18H04352.  SZ is supported in part by ECS Grant 26300316 and GRF Grant 16301917 and 16304418 from the Research Grants Council of Hong Kong.
\end{acknowledgments}



\begin{thebibliography}{99}
\bibitem{footnote3}
The following analysis can be generalized to a rotating open string with Neumann boundary conditions in a straightforward manner. It is achieved simply by multiplying a factor of $1/2$ to the Lagrangian, energy and spin. The final results are equivalent to the $N$-folded closed string ones with $N=1/2$.
\bibitem{footnote4}
The scale $\Lambda$ would depend on the model of de Sitter construction. For example, if we identify $\Lambda$ with the Planck scale in higher dimensions, its value depends on details of compactification. In general we obtain a weaker constraint for a lower $\Lambda$ with a fixed $M_s$.
\end{thebibliography}
\end{document}